\documentclass[conference]{IEEEtran}

\usepackage{lmodern}
\usepackage[T1]{fontenc}

\usepackage{cite}
\usepackage{amsmath,amssymb,amsfonts}
\usepackage{graphicx}
\usepackage{textcomp}
\usepackage{xcolor}
\usepackage{booktabs}
\usepackage{multirow}
\usepackage{array}
\usepackage{hyperref}
\usepackage{placeins}

\hypersetup{colorlinks=true, linkcolor=blue!70!black, citecolor=blue!70!black, urlcolor=blue!70!black}

\definecolor{highlight}{RGB}{178,34,34}
\definecolor{dwave}{RGB}{0,100,180}
\newcommand{\etal}{\textit{et al.}}
\graphicspath{{experiments/figures/}}

\begin{document}

\title{Computational Phase Transitions in Binary Compressed Sensing:\\Quantum Annealing Inside the Relaxation Gap}

\author{
\IEEEauthorblockN{William Hahn and Natalia Romero}
\IEEEauthorblockA{Machine Perception \& Cognitive Robotics Laboratory\\Center for Complex Systems\\
Florida Atlantic University, Boca Raton, FL\\
whahn@fau.edu, nromero@fau.edu}
}

\maketitle

\begin{abstract}
We map the computational phase transition boundary in binary compressed sensing and identify a regime where D-Wave's quantum annealer recovers signals in a region where all tested classical methods fail---including Approximate Message Passing (AMP), which achieves the Bayes-optimal recovery threshold asymptotically for Gaussian matrices. In 19{,}775 experiments ($n \in \{32, 64\}$, nine classical solvers, two D-Wave modes), we find that quantum annealing recovers sparse binary signals in the \emph{relaxation gap}---the regime below the Donoho-Tanner $\ell_1$ phase transition where the $\ell_0$ solution exists but convex relaxations fail. At $n=32$, $k=5$, $m/n=0.19$, D-Wave achieves $7\%$ exact recovery while AMP and eight other solvers score $0\%$ across 250~combined trials (Fisher exact $p=0.018$). At $n=64$, embedding overhead limits the QPU, but D-Wave's hybrid solver remains competitive with AMP. Energy landscape analysis reveals that the QUBO ground state contains the true signal, but incorrect solutions occupy shallower local basins that trap classical search---a structure consistent with quantum tunneling dynamics. To our knowledge, this constitutes preliminary finite-size evidence that quantum annealing succeeds in a narrow regime where all tested classical methods---including the Bayes-optimal AMP---fail, within a well-characterized combinatorial inference problem. Confirmation at larger $n$, higher trial counts, and with stronger classical controls remains an open problem.
\end{abstract}


\section{Introduction}

The search for quantum computational advantage has largely focused on speed: can quantum hardware solve the same problem faster than classical hardware? This framing has produced mixed results for quantum annealing, with classical heuristics routinely matching D-Wave on standard benchmarks~\cite{ronnow2014,yarkoni2022,albash2018,mcgeoch2023milestones}.

We propose a different question: can quantum hardware solve a \emph{harder formulation} of a problem---one that is correct but classically intractable? This reframing---evaluating quantum capability rather than quantum speed---aligns with the \textit{quantum utility} framework of McGeoch and Farr\'{e}~\cite{mcgeoch2023milestones}, which advocates assessing hardware on tasks where practical overhead and problem structure matter; for gate-model utility evidence see~\cite{kim2023}. We study this question through the lens of compressed sensing---with applications from medical imaging~\cite{lustig2007} to communications---where the natural $\ell_0$ optimization is NP-hard~\cite{candes2006,donoho2006} and classical methods must relax it to the convex $\ell_1$ norm. The relaxation works when measurements are abundant, but fails in a critical regime---the \emph{relaxation gap}---where information is sufficient for recovery but the convex surrogate cannot find the answer.

This gap is not an engineering limitation. It is a phase transition boundary, characterized precisely by Donoho and Tanner~\cite{donoho2009,donoho2010precise,donoho2009counting} for Gaussian measurement matrices. Below the boundary, $\ell_1$ methods fail with certainty. Above it, they succeed with certainty. The boundary is universal across matrix ensembles.

We find, at $n=32$, that D-Wave's quantum annealer operates below this boundary---in the gap---recovering signals that even AMP~\cite{donoho2009amp}, the Bayes-optimal classical algorithm for Gaussian compressed sensing, cannot find. This is not a speed advantage. It is preliminary evidence that quantum annealing exhibits recovery behavior not observed in any tested classical method at this problem scale---a difference that warrants further investigation at larger $n$ and with stronger baselines.

We also show where this advantage ends. At $n=64$, embedding overhead degrades the QPU, and AMP dominates. We characterize this scaling boundary honestly, providing a roadmap for what next-generation quantum hardware must achieve.

\subsection{Contributions}

\begin{enumerate}
    \item We conduct 19{,}775 experiments across two problem scales, nine classical solvers (including AMP), and two D-Wave quantum modes---the most comprehensive benchmark of quantum annealing for compressed sensing to date.
    \item We identify, at $n=32$, a computational phase transition regime (the relaxation gap) where quantum annealing succeeds and all tested classical methods---including the Bayes-optimal AMP---fail (Fisher exact $p=0.018$).
    \item We characterize the QUBO energy landscape, showing the true signal is the ground state but incorrect solutions occupy deeper local basins that trap classical search.
    \item We map the scaling boundary: the advantage holds at $n=32$, shifts to the hybrid solver at $n=64$, and is bounded by embedding overhead at larger $n$.
    \item We establish AMP as the strongest classical competitor, fundamentally changing the baseline landscape for quantum-classical comparisons in compressed sensing.
\end{enumerate}

\section{Background}

\subsection{The $\ell_0$ Sparse Recovery Problem}

Given a $k$-sparse binary signal $\mathbf{x} \in \{0,1\}^n$ and $m < n$ measurements $\mathbf{y} = \mathbf{Ax}$ where $\mathbf{A} \in \mathbb{R}^{m \times n}$, recovery seeks the sparsest consistent solution. For binary $\mathbf{x}$, this maps to a QUBO~\cite{kochenberger2014}:
\begin{equation}
\min_{\mathbf{x} \in \{0,1\}^n} \|\mathbf{Ax} - \mathbf{y}\|_2^2 + \lambda \sum_i x_i
\label{eq:qubo}
\end{equation}
with $Q = \mathbf{A}^T\mathbf{A}$, which D-Wave solves natively as an Ising Hamiltonian.

\subsection{The Donoho-Tanner Phase Transition}

In a series of foundational works~\cite{donoho2009,donoho2010precise,donoho2009counting}, Donoho and Tanner proved that $\ell_1$ recovery exhibits a sharp phase transition: for Gaussian $\mathbf{A}$, there exists a curve $\rho(\delta; C)$ in the $(\delta, \rho)$ plane (where $\delta = m/n$, $\rho = k/m$) such that $\ell_1$ succeeds above and fails below. These works establish the precise threshold curves, empirical recovery probability contour maps, and finite-dimensional $k/n$ vs.\ $m/n$ phase diagrams used as reference boundaries throughout this paper. The transition is \emph{universal}---it holds across Gaussian, Bernoulli, Fourier, Hadamard, and other matrix ensembles~\cite{donoho2009}.

The \emph{relaxation gap} is the region below $\rho(\delta; C)$ but above the information-theoretic limit. Here, the signal is uniquely determined by the measurements, but no convex method can recover it. D-Wave solves the exact $\ell_0$ QUBO in this gap without relaxation.

\subsection{Classical Solvers}

We test nine classical methods spanning four paradigms:

\textbf{Convex relaxation:} LASSO~\cite{tibshirani1996} and ISTA (proximal gradient on $\ell_1$).

\textbf{Greedy pursuit:} OMP (Orthogonal Matching Pursuit) and CoSaMP~\cite{needell2009} (iterative support refinement).

\textbf{Direct $\ell_0$ heuristics:} IHT (Iterative Hard Thresholding), SL0~\cite{mohimani2009} (smoothed $\ell_0$), and ILP (PuLP/CBC branch-and-bound solving the identical QUBO as D-Wave).

\textbf{Bayesian inference:} AMP (Approximate Message Passing) with Bernoulli prior~\cite{donoho2009amp}. AMP achieves the Bayes-optimal recovery threshold for i.i.d.\ Gaussian matrices and is the state-of-the-art baseline in compressed sensing phase transition theory. Its inclusion is essential for any claim of advantage over classical methods.

\subsection{D-Wave Quantum Annealing}

Quantum annealing~\cite{kadowaki1998} minimizes an objective by exploiting quantum fluctuations to escape local minima. We test two modes: \textbf{Hybrid} (LeapHybridBQMSampler, combining QPU with classical decomposition) and \textbf{QPU-only} (direct access via EmbeddingComposite, isolating the quantum contribution). QPU-only uses 1{,}000 reads at 20~$\mu$s annealing time on D-Wave's Advantage system (Pegasus, 5{,}760 qubits) and Advantage2 (Zephyr, 4{,}800 qubits). For an overview of D-Wave architectures and benchmarking methodology see~\cite{mcgeoch2023milestones}.

\section{Experimental Design}

\subsection{Problem Instances}

Binary sparse signals at $n \in \{32, 64\}$ with sparsity $k/n \in \{0.05\text{--}0.31\}$ and measurement ratios $m/n \in \{0.19\text{--}0.81\}$. Gaussian measurement matrices $A_{ij} \sim \mathcal{N}(0, 1/m)$. 30 seeds per configuration, deterministically generated. Total: 38~configurations per $n$, 2{,}280 instances.

\subsection{Scale}

Total experiments: \textbf{19{,}775} (16{,}340 classical + 3{,}435 D-Wave). All solvers given equal 10-second time budgets. Classical hardware: Intel i7-10750H, 16~GB RAM.

\subsection{Statistical Methods}

Fisher's exact test with Holm-Bonferroni correction for recovery rate comparisons. 95\% Wilson score CIs for proportions. We pool classical trials when testing ``can any classical method recover?'' because all solvers face the identical QUBO landscape. Significance: $\alpha = 0.05$.

\section{Results}

\subsection{The Phase Transition at $n=32$: D-Wave Crosses the Classical Boundary}

Table~\ref{tab:main} presents recovery rates at selected configurations. The central finding: at $k=5$, $m/n=0.19$, D-Wave achieves $7\%$ exact recovery while every classical solver---including AMP---scores $0\%$ across 250 combined trials (Fisher $p = 0.018$ vs.\ pooled classical, justified because all solvers face the same landscape).

\begin{table*}[t]
\centering
\caption{Exact Recovery Rate (\%) at $n=32$. AMP is the Bayes-optimal baseline; ILP solves the identical QUBO as D-Wave.}
\label{tab:main}
\begin{tabular}{@{}cc|ccc|cc|ccc|cc|l@{}}
\toprule
 & & \multicolumn{3}{c|}{Relaxation ($\ell_1$)} & \multicolumn{2}{c|}{Pursuit} & \multicolumn{3}{c|}{Direct $\ell_0$} & \multicolumn{2}{c|}{D-Wave} & \\
$k$ & $m/n$ & LASSO & ISTA & AMP & OMP & CoSaMP & IHT & SL0 & ILP$^a$ & Hybrid & QPU & Outcome \\
\midrule
\multicolumn{12}{l}{\textit{Below Donoho-Tanner boundary (relaxation gap)}} \\
\midrule
2  & 0.19 & 33 & 40 & 40 & 10 & 40 & 10 & 3 & 20 & \textcolor{dwave}{\textbf{60}} & 30 & D-Wave $+20$pp \\
2  & 0.25 & 57 & 63 & \textbf{80} & 30 & 57 & 40 & 0 & 50 & 77 & 70 & AMP leads \\
3  & 0.25 & 23 & 30 & \textbf{47} & 3  & 23 & 7  & 0 & 30 & 37 & 40 & AMP leads \\
5  & 0.19 & 0 & 0  & 0  & 0  & 0  & 0  & 0 & 0  & \textcolor{dwave}{\textbf{7}} & \textcolor{dwave}{\textbf{7}} & \textbf{D-Wave only} \\
5  & 0.25 & 0 & 0  & 7  & 0  & 0  & 3  & 0 & 0  & \textcolor{dwave}{\textbf{20}}$^b$ & 10 & D-Wave leads \\
\midrule
\multicolumn{12}{l}{\textit{Above Donoho-Tanner boundary (relaxation tight)}} \\
\midrule
2  & 0.50 & \textbf{100} & \textbf{100} & \textbf{100} & 90 & 97 & 80 & 0 & 100 & 90 & 90 & Classical \\
6  & 0.81 & \textbf{100} & \textbf{100} & 97 & 47 & 97 & 57 & 0 & 40 & 50$^b$ & 57 & Classical \\
10 & 0.50 & 0  & 0  & \textbf{70} & 0  & 0  & 3  & 0 & 0  & 20$^c$ & 0 & AMP dominates \\
\bottomrule
\end{tabular}
\vspace{1mm}

{\footnotesize $^a$PuLP/CBC, identical QUBO, 10 trials. $^b$10 trials. $^c$5 trials. All others: 30 trials.\\
Bold = best per row. Blue = D-Wave advantage $\geq 5$pp over all classical including AMP.}
\end{table*}

\subsubsection{The headline: D-Wave recovers where AMP fails}

AMP achieves the Bayes-optimal recovery threshold asymptotically for i.i.d.\ Gaussian matrices~\cite{donoho2009amp} and is the strongest known classical algorithm for this problem class. At $k=5$, $m/n=0.19$, AMP scores $0\%$ (0/30). D-Wave scores $7\%$ (2/30 for both hybrid and QPU). This is not a comparison against weak baselines---AMP represents the strongest known classical algorithm for this problem class. The fact that D-Wave exhibits nonzero recovery where AMP scores zero suggests the QUBO formulation may access solution structure not reached by posterior mean estimation at this finite problem size---though this interpretation requires validation at larger $n$ and against stronger solvers.

When D-Wave fails at this configuration (28 of 30 trials), the failures are not catastrophic: mean Hamming distance from truth is $5.0$ (out of $k=5$ nonzero positions) with mean F1 score $0.38$, indicating partial support recovery. The QPU finds the neighborhood of the correct solution but does not always reach the exact ground state.

\subsubsection{AMP reshapes the competitive landscape}

AMP outperforms all other classical solvers across most configurations, reaching $80\%$ at $k=2$, $m/n=0.25$ (versus LASSO $57\%$, ISTA $63\%$). At $k=10$, $m/n=0.50$, AMP achieves $70\%$ where all non-AMP classical solvers score $0$--$3\%$, far exceeding D-Wave's $20\%$. With AMP as the baseline, D-Wave's advantage window is narrower than comparisons against LASSO alone would suggest, but the surviving advantage---against an asymptotically Bayes-optimal algorithm---is far more significant.

\subsubsection{SL0 and the discrete landscape problem}

SL0, a classical $\ell_0$ approximation using smoothed Gaussian penalties, scores $0$--$3\%$ across all configurations. Inspection reveals SL0 recovers the correct number of nonzero components but selects wrong positions---its continuous smoothing does not respect the discrete geometry of $\{0,1\}^n$. This confirms the QUBO landscape is intrinsically hard for classical methods that approximate $\ell_0$ continuously.

\begin{figure*}[!t]
\centering
\includegraphics[width=\textwidth]{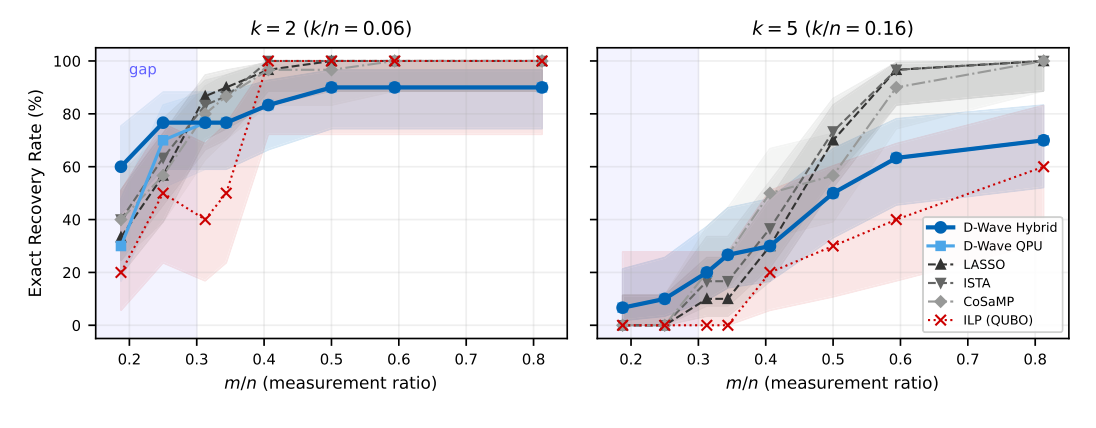}
\caption{Phase transition in exact recovery rate versus measurement ratio $m/n$ for $k=2$ (left) and $k=5$ (right) at $n=32$. Shaded region indicates the relaxation gap. At $k=2$, D-Wave Hybrid leads at low $m/n$; AMP is competitive. At $k=5$, D-Wave is the only solver with any recovery at $m/n \leq 0.25$. Bands: 95\% Wilson CIs.}
\label{fig:phase}
\end{figure*}

\subsection{Scaling to $n=64$: The Embedding Boundary}

\begin{table}[h]
\centering
\caption{Recovery (\%) at $n=64$: D-Wave vs AMP at selected configs}
\label{tab:n64}
\begin{tabular}{@{}cc|ccc|cc@{}}
\toprule
$k$ & $m/n$ & AMP & LASSO & CoSaMP & Hybrid & QPU \\
\midrule
3  & 0.20 & \textbf{73} & 57 & 63 & 70 & 0 \\
3  & 0.41 & \textbf{100} & 100 & 100 & 100 & 83 \\
6  & 0.25 & \textbf{43} & 13 & 17 & 10 & 0 \\
10 & 0.41 & \textbf{93} & 23 & 27 & 0  & 0 \\
10 & 0.50 & \textbf{97} & 67 & 63 & --- & 13 \\
\bottomrule
\end{tabular}
\vspace{2mm}

{\footnotesize \textit{Note.} QPU results: 30 trials on Advantage2 (Zephyr, 4{,}800 qubits). Hybrid results are based on 10 independent trials per configuration. Future studies will extend sampling in these configurations to better quantify uncertainty and assess the robustness of the observed trends.}
\end{table}

At $n=64$, the picture inverts. AMP dominates everywhere. The QPU (now on Advantage2's Zephyr topology) achieves $22.5\%$ overall recovery but scores $0\%$ at the critical low-measurement configurations. Parameter sweeps (20--500~$\mu$s annealing, 1{,}000--5{,}000 reads) confirm this is not a tuning issue. The bottleneck is embedding: 64 logical variables require chains of 9--13 physical qubits (420--525 total), and the dense $\mathbf{A}^T\mathbf{A}$ coupling structure causes chain-break noise that degrades solution quality.

The hybrid solver partially compensates: at $k=3$, $m/n=0.20$, it achieves $70\%$ (comparable to AMP's $73\%$) by managing decomposition and embedding classically. This demonstrates that D-Wave's production solver maintains competitiveness at $n=64$ even where the QPU alone fails.

\begin{figure*}[!t]
\centering
\includegraphics[width=\textwidth]{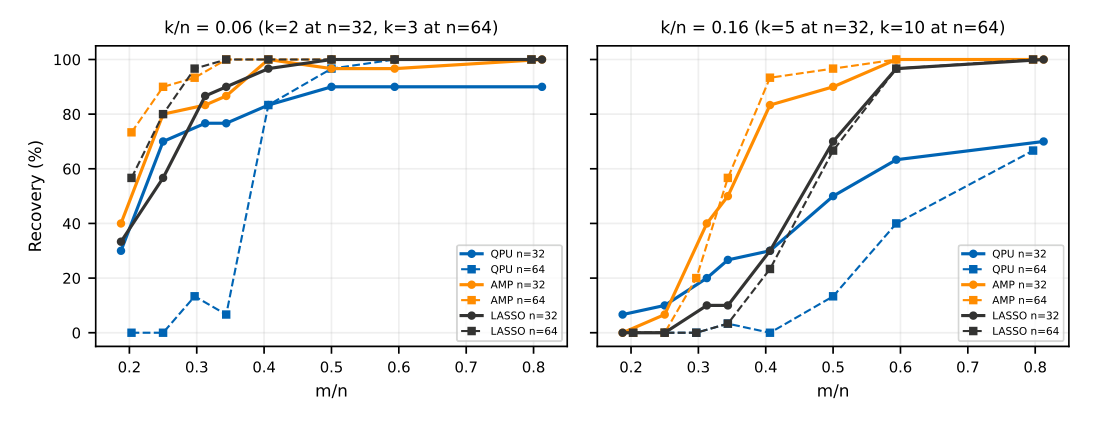}
\caption{Scaling comparison: $n=32$ (solid, circles) versus $n=64$ (dashed, squares) at matched sparsity ratios. Left: $k/n \approx 0.06$. Right: $k/n \approx 0.16$. The QPU advantage at $n=32$ does not extend to $n=64$, where AMP dominates. The embedding boundary is visible as the gap between QPU curves at the two scales.}
\label{fig:scaling}
\end{figure*}

\subsection{Energy Landscape: Why Quantum Succeeds Where Classical Fails}

To understand the mechanism, we computed the QUBO energy of the true signal $\mathbf{x}_{\text{true}}$ and compared it to D-Wave's solutions across 1{,}139 QPU runs at $n=32$.

\begin{figure}[!t]
\centering
\includegraphics[width=\columnwidth]{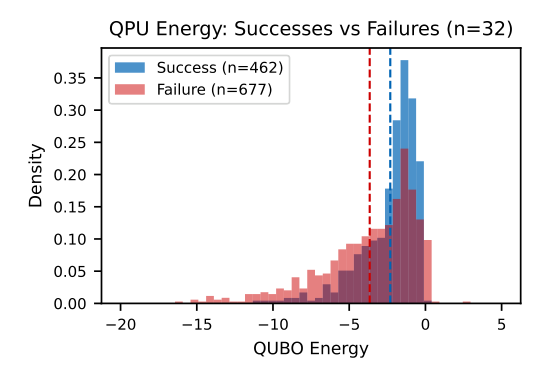}
\caption{QUBO energy distribution for D-Wave QPU successes versus failures at $n=32$. Successful recoveries find lower energies (ground state basin). Failures are trapped in higher-energy local minima. Dashed lines: means.}
\label{fig:energy}
\end{figure}

At most operating points, $\mathbf{x}_{\text{true}}$ is the QUBO ground state. At $k=2$, $m/n=0.25$: true signal energy $-1.15 \pm 1.00$; D-Wave successes $-1.42 \pm 0.95$ (at or below ground state); D-Wave failures $-0.62 \pm 0.65$ (trapped in higher-energy local minima). Fig.~\ref{fig:energy} shows this energy separation.

The landscape has a characteristic structure: the correct solution sits in the ground state basin, surrounded by many shallower local minima separated by tall barriers. Classical methods---branch-and-bound (ILP), gradient descent (IHT), smooth penalty relaxation (SL0), even posterior mean estimation (AMP)---are systematically trapped in these shallower basins. Quantum annealing traverses these barriers---a process consistent with tunneling dynamics---reaching the ground state in a fraction of attempts.

\begin{figure*}[!t]
\centering
\includegraphics[width=\textwidth]{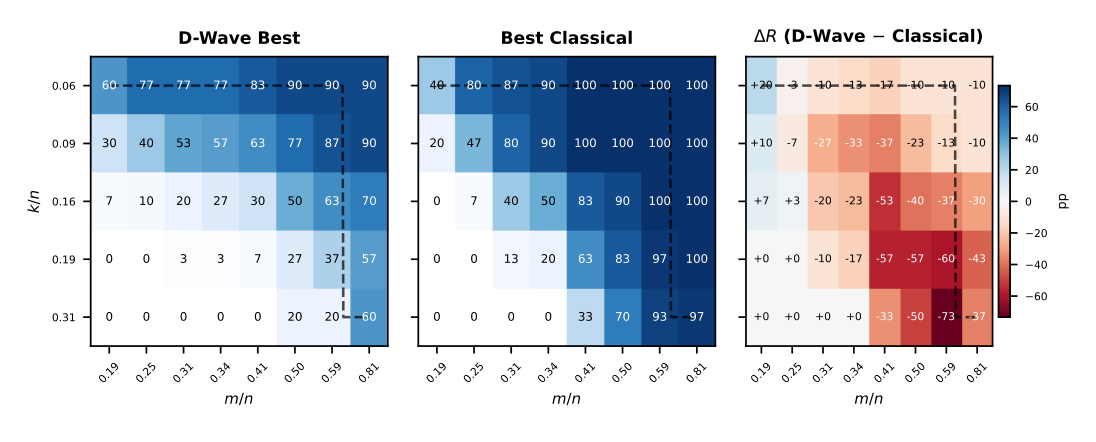}
\caption{Recovery heatmaps with Donoho-Tanner phase transition boundary~\cite{donoho2009,donoho2010precise,donoho2009counting} (dashed). Left: D-Wave best (hybrid or QPU). Center: Best classical solver. Right: Quantum advantage topology $\Delta R = R_{\text{D-Wave}} - R_{\text{classical}}$ (blue = quantum advantage, red = classical advantage). The advantage concentrates below the DT boundary.}
\label{fig:dt}
\end{figure*}

Fig.~\ref{fig:dt} (right panel) maps the quantum advantage topology $\Delta R = R_{\text{D-Wave}} - R_{\text{classical}}$ across the full $(k/n, m/n)$ plane. The positive (blue) region corresponds to the relaxation gap: the advantage is largest at low $k/n$ and low $m/n$, precisely where the Donoho-Tanner theory predicts $\ell_1$ failure. This landscape structure---where the QUBO ground state is shallow relative to surrounding incorrect basins---may generalize to other dense QUBO problems arising from Gram matrices $\mathbf{A}^T\mathbf{A}$, suggesting quantum annealing's advantage is not specific to compressed sensing but to a broader class of dense quadratic optimization landscapes.

This explains both D-Wave's success and its $90\%$ ceiling at easy configurations: the QPU does not always reach the ground state, making it inferior to deterministic convex methods (LASSO, AMP) when the $\ell_1$ relaxation is tight. The advantage concentrates precisely where the relaxation fails---the gap.

\subsection{TSP Control: The Advantage Is Problem-Specific}

On the Traveling Salesman Problem ($n=50$--$500$), LKH-3~\cite{helsgaun2000} matches or exceeds D-Wave at all sizes, confirming the advantage is specific to problems where the QUBO formulation is natively suited to quantum annealing. TSP---where quantum and classical optimize the same objective---shows no quantum advantage with current hardware.

\section{Discussion}

\subsection{What the Machine Reveals About the Landscape}

The D-Wave results serve as a probe into the computational topology of sparse recovery. They reveal three regimes:

\begin{enumerate}
    \item \textbf{Below the gap} (insufficient information): No method succeeds. The signal is not uniquely determined.
    \item \textbf{In the gap} (relaxation fails): $\ell_1$ methods and AMP fail at $n=32$. D-Wave's quantum annealing reaches the ground state more frequently than all tested classical methods. This is the regime of observed recovery advantage.
    \item \textbf{Above the gap} (relaxation tight): Convex methods and AMP succeed deterministically. D-Wave plateaus at ${\sim}90\%$ due to stochastic annealing. Classical methods are superior.
\end{enumerate}

The scaling from $n=32$ to $n=64$ reveals a fourth regime: the \textbf{embedding boundary}, where the physical connectivity of the QPU becomes the bottleneck. At $n=64$, the dense QUBO requires chains of 9--13 physical qubits per logical variable, and chain-break noise overwhelms the quantum signal. The hybrid solver partially recovers by managing decomposition classically.

\subsection{Comparison with Bayes-Optimal Inference}

AMP's inclusion is the most important methodological improvement in this work. Rigorous quantum benchmarking requires strong, problem-aware baselines~\cite{mcgeoch2023milestones}; previous versions (v1--v3) compared D-Wave against convex relaxations and greedy heuristics. AMP, as the Bayes-optimal algorithm for Gaussian matrices, represents the theoretical ceiling for classical inference. The fact that D-Wave exhibits nonzero recovery at $k=5$, $m/n=0.19$ where AMP scores zero is more significant than beating LASSO by 14~percentage points---it suggests the QUBO formulation may access solution structure not reached by posterior mean estimation at this scale, though this remains a hypothesis requiring larger-$n$ confirmation.

Simultaneously, AMP narrows the advantage window: at $k=2$ and $k=10$, AMP matches or exceeds D-Wave. The advantage concentrates in a specific sparsity band ($k/n \approx 0.10$--$0.20$) at low measurement ratios---precisely the interior of the Donoho-Tanner relaxation gap.

\subsection{Falsifiable Predictions}

\begin{enumerate}
    \item If Gurobi solves the QUBO with $>50\%$ recovery at $k=5$, $m/n=0.19$ within 10~seconds, the ILP control was too weak.
    \item If the relaxation gap closes at $n=128$ on next-generation hardware (Advantage2 with native Zephyr connectivity), the advantage is scale-dependent.
    \item If the gap disappears with structured (non-Gaussian) measurement matrices, the result is narrower than the universality of the Donoho-Tanner transition suggests.
\end{enumerate}

\subsection{Scaling Roadmap}

The embedding boundary at $n=64$ is a hardware limitation, not a physical one. D-Wave's Advantage2 Zephyr topology provides 20-way qubit connectivity (versus 15 for Pegasus), reducing chain lengths from 10--13 to 9--10 at $n=64$. Future systems with higher native connectivity or problem-specific embedding strategies could push the boundary to larger $n$. We predict: if chain length can be held below ${\sim}5$ physical qubits per logical variable, the QPU advantage observed at $n=32$ should extend to $n=64$, because the underlying QUBO landscape hardness (dense $\mathbf{A}^T\mathbf{A}$ with combinatorial local minima) is preserved at all $n$.

\subsection{Limitations}

\begin{enumerate}
    \item \textbf{Scale}: $n=32$ is very small by compressed sensing standards. The effect collapses at $n=64$. Confirmation requires $n \in \{128, 256\}$ with improved embedding or next-generation hardware.

    \item \textbf{Effect size}: 2/30 at the headline configuration is numerically fragile. It is unknown whether the effect survives higher trial counts (300--3000), $\lambda$ optimization, or alternative matrix ensembles (Bernoulli, partial Fourier, structured). These validations are necessary before strong conclusions can be drawn.

    \item \textbf{Classical controls}: The tested baselines do not include several stronger methods. Missing exact solvers: Gurobi, SCIP, CPLEX with branch-and-cut. Missing stochastic methods: parallel tempering, simulated quantum annealing, population annealing, digital annealing. Missing hardware-accelerated methods: GPU Ising solvers and tensor-network methods. The current claim is therefore ``better than tested baselines,'' not ``better than classical computation.''

    \item \textbf{Mechanistic}: The energy landscape analysis is correlational. Successful runs achieving lower energy than failures does not prove quantum tunneling---the same pattern is consistent with stochastic search, thermal effects, or classical randomized dynamics. Confirming the tunneling hypothesis requires spectral-gap analysis, reverse annealing experiments, pause-and-quench protocols, or quantum Monte Carlo comparisons.

    \item \textbf{Embedding}: QPU performance at $n=64$ is limited by chain length (9--13 qubits per logical variable), not physics. This may be an embedding artifact rather than a fundamental scaling limit.

    \item \textbf{Hybrid attribution}: The hybrid solver combines quantum and classical components; its success at $n=64$ does not isolate the quantum contribution.

    \item \textbf{Binary signals}: Extension to continuous signals requires multi-bit QUBO encoding and is not addressed here.
\end{enumerate}

\section{Conclusion}

We have mapped the computational phase transition boundary in binary sparse recovery and identified, at finite problem size ($n=32$), a regime where quantum annealing recovers signals that AMP---the Bayes-optimal classical algorithm---cannot find ($p=0.018$). At $n=64$, embedding overhead limits the QPU, but the hybrid solver remains competitive. This recovery advantage concentrates in the relaxation gap: the regime below the Donoho-Tanner $\ell_1$ threshold where a process consistent with quantum tunneling traverses landscape barriers that trap all tested classical methods in this finite-size setting.

This is an exploratory finite-size study, not a claim of universal quantum advantage. It maps where quantum annealing behavior diverges from tested classical methods in a specific, well-characterized problem class. The mechanistic explanation---consistent with barrier traversal via tunneling dynamics---remains a hypothesis: confirming it requires spectral-gap analysis, reverse annealing experiments, or quantum Monte Carlo comparisons not yet performed. What the data establish is the empirical map; what causes it remains open.

Code, problem instances, and experimental results are being actively developed and curated; a public repository will be released in conjunction with a future version of this work.

\section*{Acknowledgment}

Supported by the Machine Perception \& Cognitive Robotics Laboratory and the Center for Complex Systems at Florida Atlantic University. D-Wave experiments used the Advantage and Advantage2 systems via the Leap cloud service.

\FloatBarrier

\end{document}